\documentclass[twocolumn,floatfix,showpacs,preprintnumbers,amsmath,amssymb]{revtex4}
\usepackage{graphicx}
\usepackage{subfigure}
\usepackage{amssymb}
\usepackage{bm}

\begin{document}

\title{Ring dark solitons in microcavity polariton condensates}

\author{Szu-Cheng Cheng$^{1}$}
\author{Ting-Wei Chen$^{2}$}
\email {twchen@mail.ncyu.edu.tw}
\thanks{FAX: +886-5-2717909}

\affiliation{$^{1}$Department of Optoelectric Physics, Chinese Culture University, Taipei 11114, Taiwan, R. O. C.\\
$^{2}$Department of Electrophysics, National Chiayi University, Chiayi city 60004, Taiwan, R. O. C.}

\date{\today}

\begin{abstract}
A ring dark soliton is a dark soltion occurring in higher dimensions. It is still unknown in microcavity-polariton condensates due to its instability. We find that a stable RDS cannot exist in a MPC without a defect. We then propose a way to create a stable RDS in a MPC by adding a defect with a ring structure to the system. A RDS pinned by the defect potential becomes stable. For various pump powers, we also investigate the stable regime of the RDS by tuning the strength and width of the defect potential. We conclude that the stable RDSs can be obtained.
\end{abstract}

\pacs{03.75.Kk, 47.32.-y, 71.36.+c}

\maketitle
The nonlinearities governing the behaviors of optical systems and Bose-Einstein condensates (BECs) allow for the excitations such as spatial bright and dark solitons \cite{NOL1,NOL2,NOL3,NOL4}. For the dark solitons, i.e., the intensity dips with a phase jump across the intensity minimum on a uniform background, the stability is known to hold only in the one-dimensional geometry. In the two-dimensional (2D) geometry, dark solitons, in the form of stripes, are prone to the transverse modulation instability or the so-called snake instability \cite{Denschlag,n3,Feder,Brand}. As a result, they may bend and eventually decay towards vortices, vortex pairs or vortex rings. However, this instability can be suppressed by bending a soliton stripe to close it into an annulus of a particular length. In nonlinear optics, this idea led to the introduction of ring dark solitons (RDSs) \cite{Kivshar}, whose properties have been studied both in theory \cite{Dreischuh1,Frantzeskakis,Nistazakis} and in experiments \cite{Neshev,Dreischuh2}.

The RDSs have also been predicted to occur in atom BECs \cite{Theocharis}, and their dynamics were analyzed by means of the perturbation theory of dark matter-wave solitons. A more recent work, trying to stabilize RDSs in the form of a dynamically robust state of quasi-2D BECs, has been reported \cite{Wang}. However, in order to achieve the requirement of the atomic BECs, extremely low temperature is necessary to study the properties of the ring solitons. Contrary to the atomic BECs, creating microcavity-polariton condensates (MPCs) is relatively easier since MPCs have been observed experimentally at room temperature \cite{room1,room2,room3}.

MPCs have been the subject of intensive research over the past two decades because of their potential advances towards a new generation of low-threshold lasers and ultrafast optical amplifiers and switches at room temperature \cite{1}. Microcavity polaritons are bosonic quasiparticles, which arise from strong exciton-photon coupling in semiconductor microcavities. Characteristic bosonic phenomena, such as stimulated scattering \cite{2} and polariton condensation have been reported \cite{3a,3b,3c,3d}. However, polaritonic systems are intrinsically out-of-equilibrium so that continuous pumping is needed to balance the fast polariton decay and maintain a steady-state solution \cite{4,5,6,7}. Thanks to these assets and with the possibility of ultrafast imaging, L. Dominici $\textit{et al.}$ reported the first experimental observation of spontaneous emergence of polaritonic RDSs activated by a resonant exciting pulse and contoured by the emission of shock waves \cite{Dominici}. Moreover, a different kind of stationary gray ring solitons can be theoretically found for polariton condensates in suitable parametric regimes \cite{Rodrigues}. Except these two articles so far, there is a lack of literature exploring the formation and stability of RDSs in the polariton condensates.

In a one-dimensional MPC, a dark soliton subjected to a uniform MPC background is unstable and displays an abrupt decay \cite{Xue}. We showed that an unstable and decaying dark soliton in a MPC could be stabilized or pinned by the presence of a defect potential \cite{Cheng12}. A RDS is a one-dimensional dark soliton along the azimuthal direction of a 2D MPC. Due to the non-equilibrium character of MPCs, the RDS is then unstable in a uniform MPC background. In this paper, we propose a method of generating a stable RDS in a MPC subjected to an annular defect potential. The RDSs of MPCs are studied via the complex Gross-Pitaevskii equation (cGPE) coupled to the reservoir polaritons from higher momenta \cite{5} with the poariton mass, $m$, and interaction strength between polaritons, $g$. This mean-field model for non-equilibrium MPCs is a generic model of considering effects from pumping, dissipation, relaxation and interactions.

For the non-equilibrium MPCs, we treat the polaritons from higher momenta as a reservoir and employ the cGPE, governing the condensate polaritons that couples to the reservoir polaritons with density $n_{\textbf{R}}(\textbf{r},t)$, to describe the time evolution and density distribution of the condensate $\Psi (\textbf{r},t)$. In the cGPE, we add an annular defect sitting at the radial distance $r_0$ from the center of the condensate. When the pump power, $P$, is larger than the threshold power, $P_{th}$, that is, ($P>P_{th}$), the condensation of polaritons occurs. The condensate density being far away from the dip region of the RDS, $n_{c}$, is given by $n_{c}=(P_{th}/\gamma)\alpha$, where $\alpha=(P/P_{th})-1$ is called the pump parameter being the relative pumping intensity above the threshold power and $\gamma$ is the loss rate of the condensate. In the mean time, the stationary reservoir density, which is determined by the net gain being zero, is equal to the reservoir density, $n^{th}_{R}$, at the threshold pump power, i.e., $n^{th}_{R}=P_{th}/\gamma_{R}$, where $\gamma_{R}$ is the loss rate of reservoir polaritons. Let $\hbar$ be Planck's constant and $\sigma=1/(1-(4\gamma/\gamma_{R}))$. We take the length, energy and time scales in units of $\lambda =\sqrt{\hbar^2\gamma\sigma/2mgP_{{\rm{th}}}}$, $\hbar\omega_{0}=\hbar^{2}/2m\lambda ^{2}$ and $\tau=t/\omega _{0}$, respectively. Then the wave function of a MPC is written as $\psi (\bm{\rho },\tau )=\Psi (\bm{\rho },\tau)/\sqrt{n_{c}}$, and the reservoir polariton density as $n(\rho,\tau)=n_{\textbf{R}}(\bm{\rho },\tau)/n^{th}_{R}$, where $\bm{\rho }=\textbf{r}/\lambda$ and $\tau=\omega _{0}t$. Using the polar coordinate $\bm{\rho }=(\rho ,\theta )$, the condensate wave function $\psi (\bm{\rho },\tau )$ and reservoir density $n(\bm{\rho },\tau)$ satisfy the cGPE equations as below \cite{Cheng14}
\begin{equation}\label{1}
i\frac{{\partial \psi }}{{\partial \tau }} =  - \nabla _\rho ^2 \psi  + \tilde V\psi  + \frac{i}{2}\left[ {\tilde{R}n - \tilde\gamma } \right]\psi \\
 + \alpha \sigma \left| \psi  \right|^2 \psi  + (\sigma  - 1)n\psi,
 \end{equation}
\begin{equation}\label{2}
\frac{{\partial n}}{{\partial \tau }} = \tilde \gamma _R \left( {\alpha  + 1 - n} \right) - 4\tilde{R}\left( {\frac{{\alpha \sigma }}{{\sigma  - 1}}} \right) n \left |\psi  \right|^2,
\end{equation}
where $\tilde{\gamma}=\gamma/\omega_{0}$ and $\tilde{\gamma}_{R}=\gamma_{R}/\omega_{0}$. The Laplacian operator $\nabla_{\rho}^{2}$ is defined as $\nabla_{\rho}^{2}\equiv \frac{1}{\rho}\frac{\partial}{\partial\rho}(\rho\frac{\partial}{\partial\rho})+ \frac{1}{\rho^{2}}\frac{\partial^{2}}{\partial\theta^{2}}$. The scaled defect potential $\tilde{V}=V_{1}e^{-(\rho-\rho_0)^{2}/a_{1}^{2}}$ with the dimensionless potential strength $V_{1}=V_{0}/\hbar\omega_{0}$, $\rho_{0}=r_0/\lambda$ and width $a_{1}=a/\lambda$. Here $\tilde {R}=R/\omega_{0}$, where $R$ is the dimensionless amplification rate that describes the replenishment of the condensate state from the reservoir state by stimulated scattering.

The steady state of Eqs. (\ref{1}) and (\ref{2}) under a uniform pumping can be obtained by taking $\psi = \psi _0 e^{ - i\tilde \mu \tau }$ and $n = n _0$, where $\tilde \mu  = \mu /\hbar \omega _0$ is the dimensionless chemical potential of the system and $\frac{{\partial n_0}}{{\partial \tau }}=0$. The solutions of Eq. (\ref{1}) and (\ref{2}) with $\tilde{V}(\rho)\neq 0$ are very different from the uniform MPC.  In the limit $\rho\rightarrow\pm\infty$, $\psi_{0}(\rho)\rightarrow 1$ and $n_{0}\rightarrow 1$, we find that $\tilde{R}=\tilde\gamma$ on the steady state of the system and the chemical potential of the system is given by $\tilde\mu=\alpha\sigma+(\sigma-1)$. Inserting $\tilde\mu=\alpha\sigma+(\sigma-1)$ into Eq. (\ref{1}), then the stationary equations of Eqs. (\ref{1}) and (\ref{2}) become
\begin{multline}\label{3}
\nabla _\rho ^2 \psi_{0}-V(x)\psi_{0}+\alpha\sigma(1-|\psi_{0}|^{2})\psi_{0}-(\sigma-1)(n_{0}-1)\psi_{0}\\
-\frac{i\tilde\gamma}{2}(n_{0}-1)\psi_{0}=0,
\end{multline}
\begin{equation}\label{4}
n_{0}(1+\alpha|\psi_{0}|^{2})-\alpha-1=0.
\end{equation}
Using the Newton-Raphson method, we can solve Eqs. (\ref{3}) and (\ref{4}) numerically. The density distribution of a RDS is a combination of the density distributions of condensate and reservoir polaritons. Due to the non-zero density of reservoir polaritons at $\rho=\rho_0$, the total density of the system is not zero at the dip of the soliton. In other words, the dip of a RDS contains some background density from reservoir polaritons. The condensate density decreases while the density of reservoir polaritons increases inside the dip of a RDS; and the total density at the center the RDS is given by the summation of the density of reservoir and condensate polaritons.
\begin{figure}
\centering
\scalebox{.24}{\includegraphics{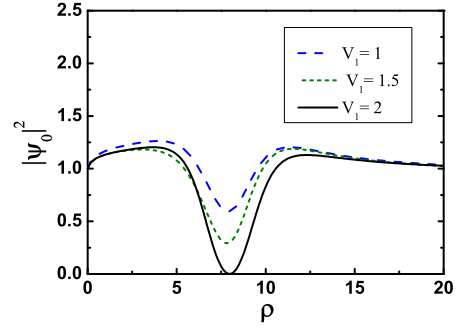}}
\caption{ (Color online) Density of condensate polaritons for various defect potential strength $V_{1}$. Parameters: $\alpha=1$, $a_{1}=2$, $\rho_0=8$.}
\end{figure}

\begin{figure}
\subfigure{\scalebox{.23}{\includegraphics{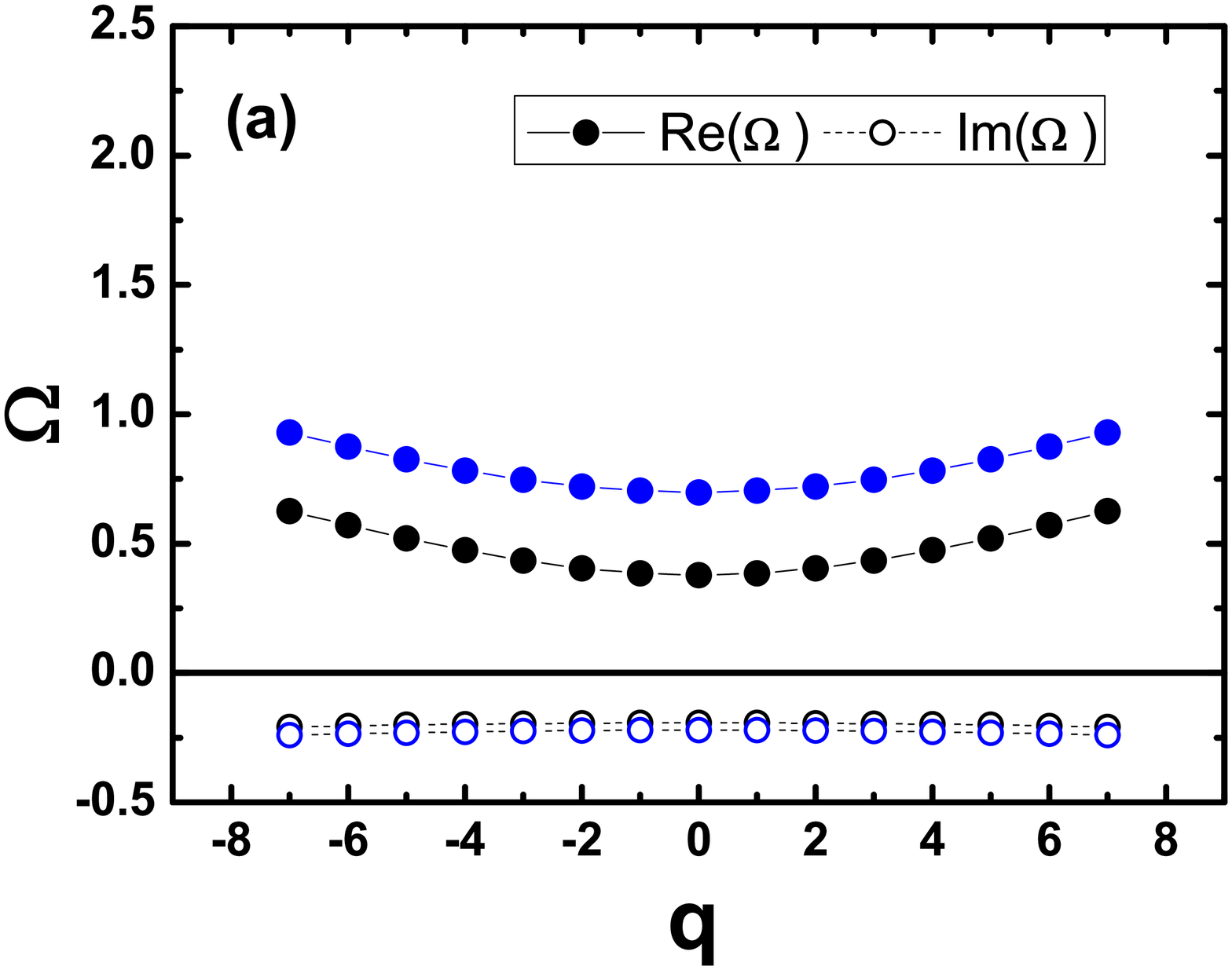}}}
\subfigure{\scalebox{.23}{\includegraphics{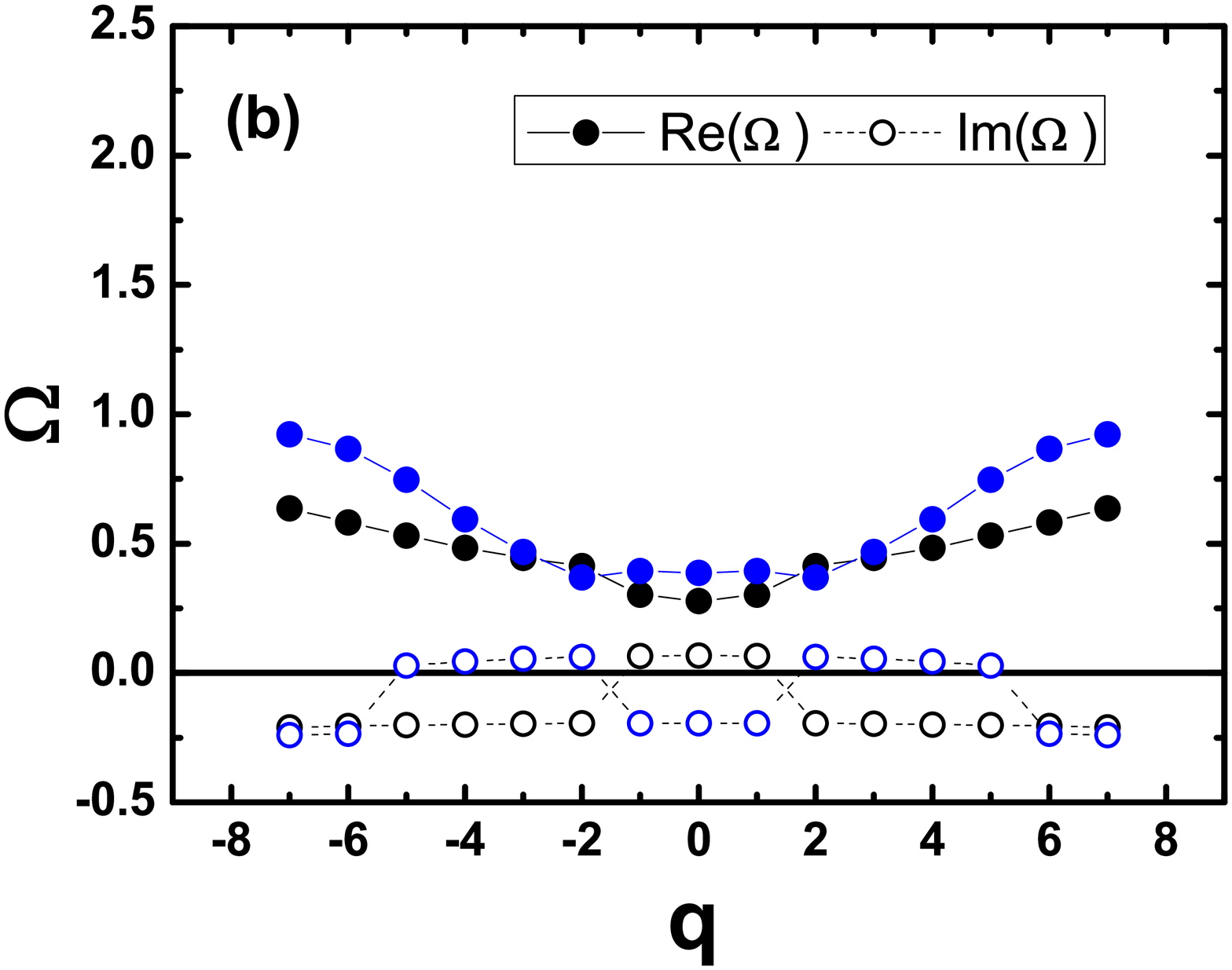}}}
\subfigure{\scalebox{.23}{\includegraphics{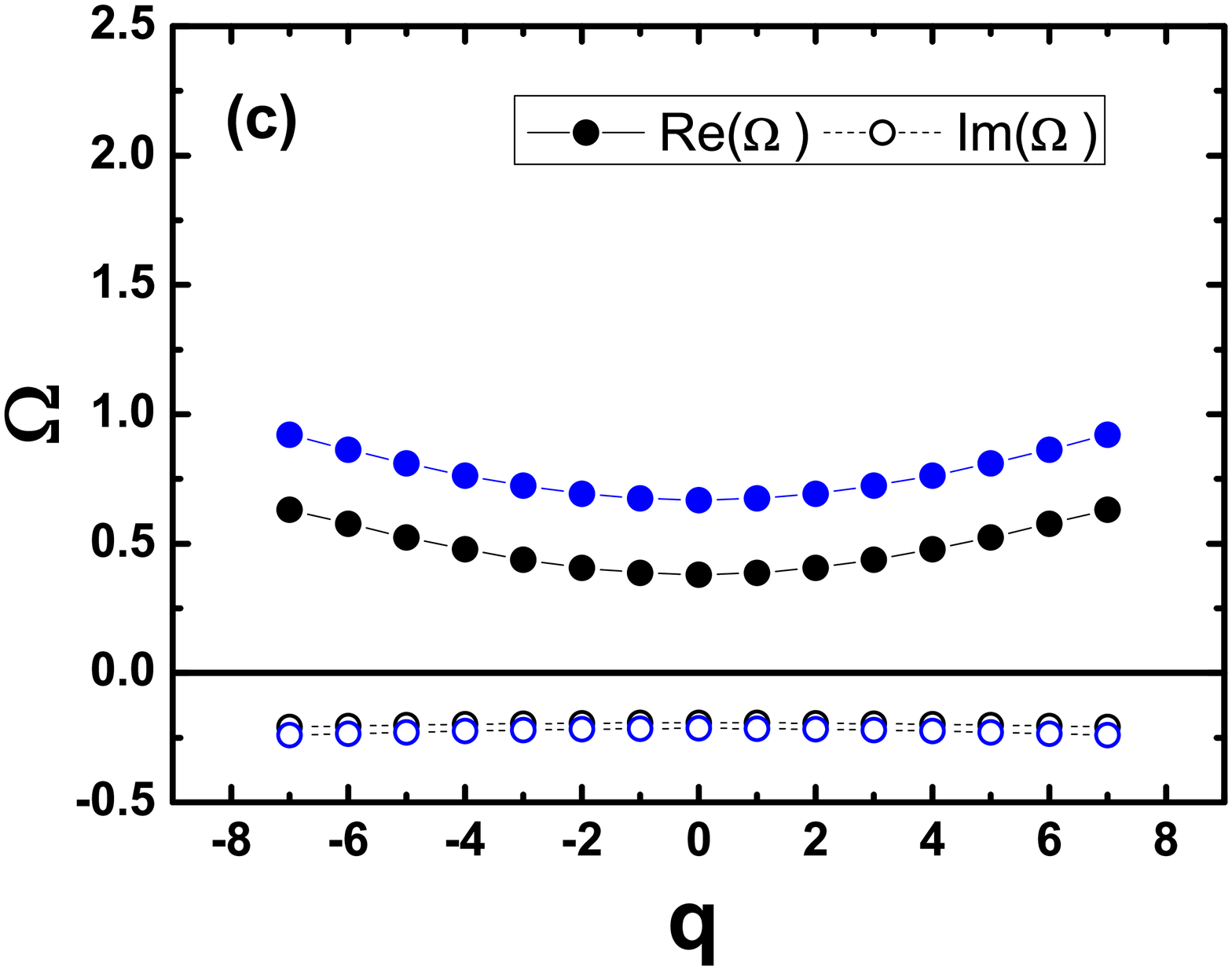}}}
\caption{ (Color online) Two low-lying excitation frequencies of steady RDSs for different defect-potential strength:(a)$V_{1}=0.3$, (b)$V_{1}=0.6$, (c)$V_{1}=1.2$. Here $\alpha=0.5$, $a_{1}=2$, $\rho_0=8$.}
\end{figure}

In Fig. 1, we show the density distribution of the condensate for the pump strength $\alpha =1$ and the defect potential width $a_{1}=2$ and various strengths $V_{1}$. For the strength $V_{1}=1$, the repulsive force repels the condensate polaritons from the dip region of the ring soliton. There are still some particles inside the dip, and the ring soliton is a ring gray soliton (RGS). However, further increasing the potential to $V_{1}=2$, the condensate density at the dip center becomes zero that leads to the formation of a RDS. The RDS and RGS have annular dark and gray soliton in the azimuthal direction, respectively. Namely, such a 2D structure consists of an inner part of the density surrounded by a condensate background that is out of phase with respect to the inner part. The two parts are separated by a region of vanishing or lower density with the size controlled by the external pumping and potential.

The excitation frequencies of steady RDSs with respect to the quasi-momentum $q$ are shown in Figure 2 for three different strengths, i.e., $V_{1}=0.3, 0.6, 1.2$ (see Supplementary for the detailed calculations on excitation frequencies). The stability of a RDS is fulfilled if Im$(\Omega)<0$, where $\Omega$ is the excitation frequency of the system. In Fig. 2(a), a gray soliton contains some particles inside the dip. These existing particles create a repulsive potential to prevent excitations from refilling particles into the dip \cite{Cheng12}. Therefore, a stabilized RGS can occur in a non-equilibrium MPC. As $V_{1}$ is an intermediate strength, there is no condensate polariton at the dip center. It is easier for excitations to redistribute the MPC density in the dip, which results in the instability of the RDS (see Fig. 2(b)). To create a stable RDS in a MPC, we can increase the defect potential strength further. Let the expulsion force from the defect become large so that the redistribution of the MPC density is prohibited. In Fig. 2(c), a RDS pinned by an even higher defect-potential strength is stable. The stability indicates that stable RGSs and RDSs in MPCs can be manipulated by the defect potential with proper strength $V_{1}$ and width $a_{1}$.
\begin{figure}
\centering
\scalebox{.24}{\includegraphics{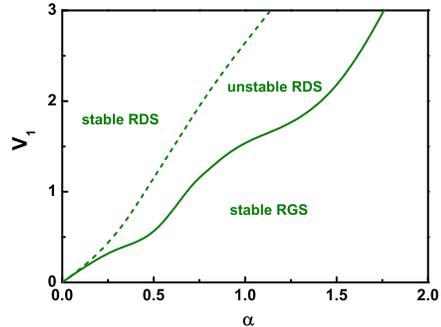}}
\caption{Phase boundary for the ring solitons in terms of $V_{1}$ and $\alpha$ under the defect potential width $a_{1}=2$.}
\end{figure}

Using the perturbation theory aforementioned, the shallow-dip RGSs in polariton condensates are ling-lived objects, that may be observed experimentally on a relevant time scale. On the contrary, the deep-rift RDSs are subject to the instability due to the combination effect from dispersion relations, nonlinearity and non-equilibrium of a MPC. In Figure 3, we show the borders between the regions of unstable RDSs, stable RDSs and that of stable RGSs in terms of various pump powers and defect-potential strengths with a fixed potential width $a_{1}=2$. The RGSs exist stably in a regime with smaller defect-potential strengths and higher pump powers. When the pump power and defect potential strength are intermediate strengths, respectively, unstable RDSs can occur in this regime. We can generate stable RDSs in a regime with lower pump powers, in which the background density of the dark soliton is low, and higher defect-potential strengths. Increasing the defect-potential strength with a fixed defect-width $a_{1}=2$, the ring soliton evolves from a stable RGS ($V_{1}\lesssim0.48$) to an unstable RDS ($0.48\gtrsim V_{1}\lesssim1.15$), and then becomes a stable RDS ($V_{1}\gtrsim1.15$). All these dynamical features are drastically different from those known RDSs in nonlinear optics and equilibrium atomic BECs. The formation of a ring soliton in a MPC is subjected to the balancing effect from dispersion relations, nonlinearity and non-equilibrium of a MPC. The non-equilibrium ring solitons show much broader physical phenomena than those ring solitons occurring in nonlinear optics and atomic BECs.

\begin{figure}
\subfigure{\scalebox{.20}{\includegraphics{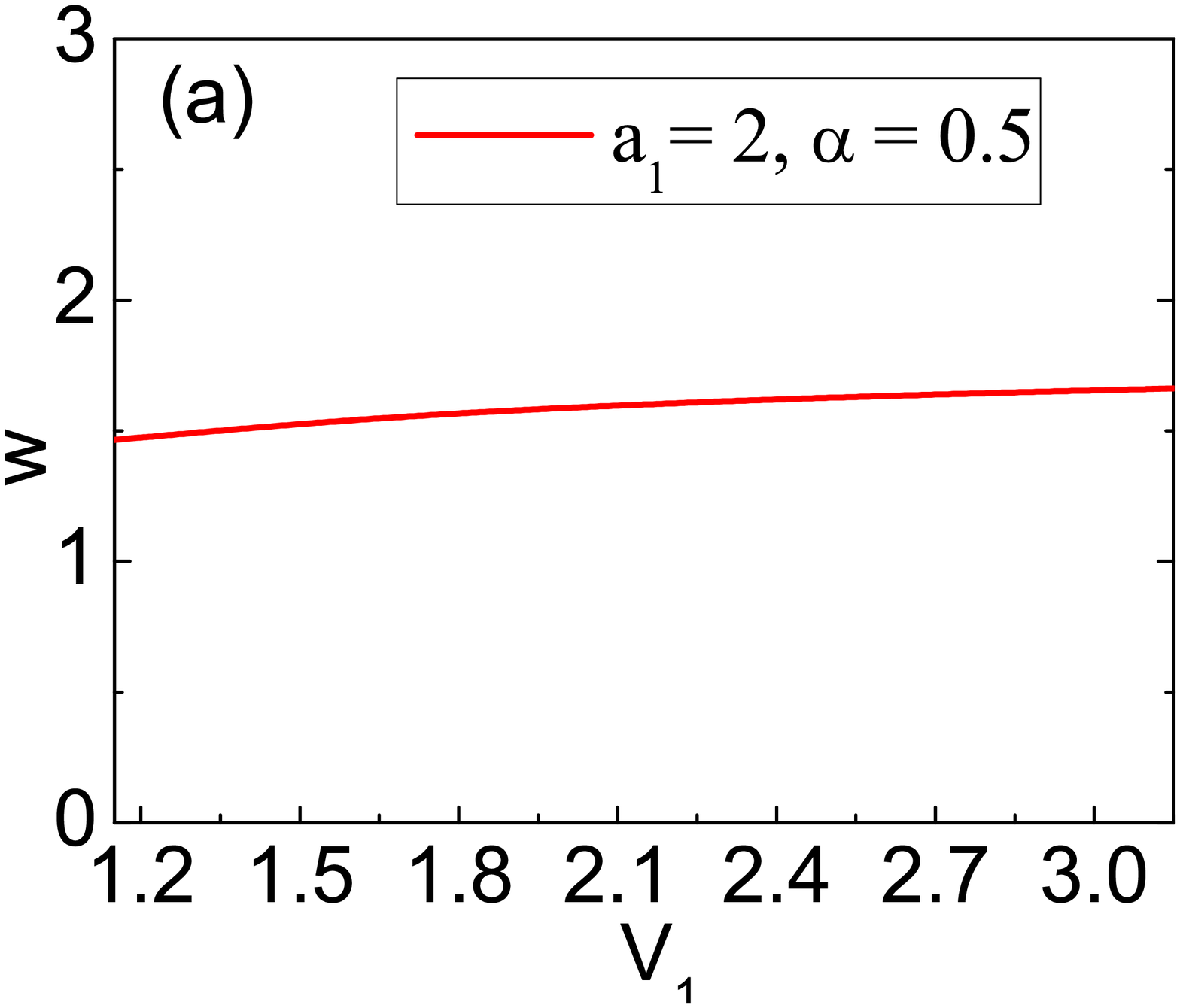}}}
\subfigure{\scalebox{.20}{\includegraphics{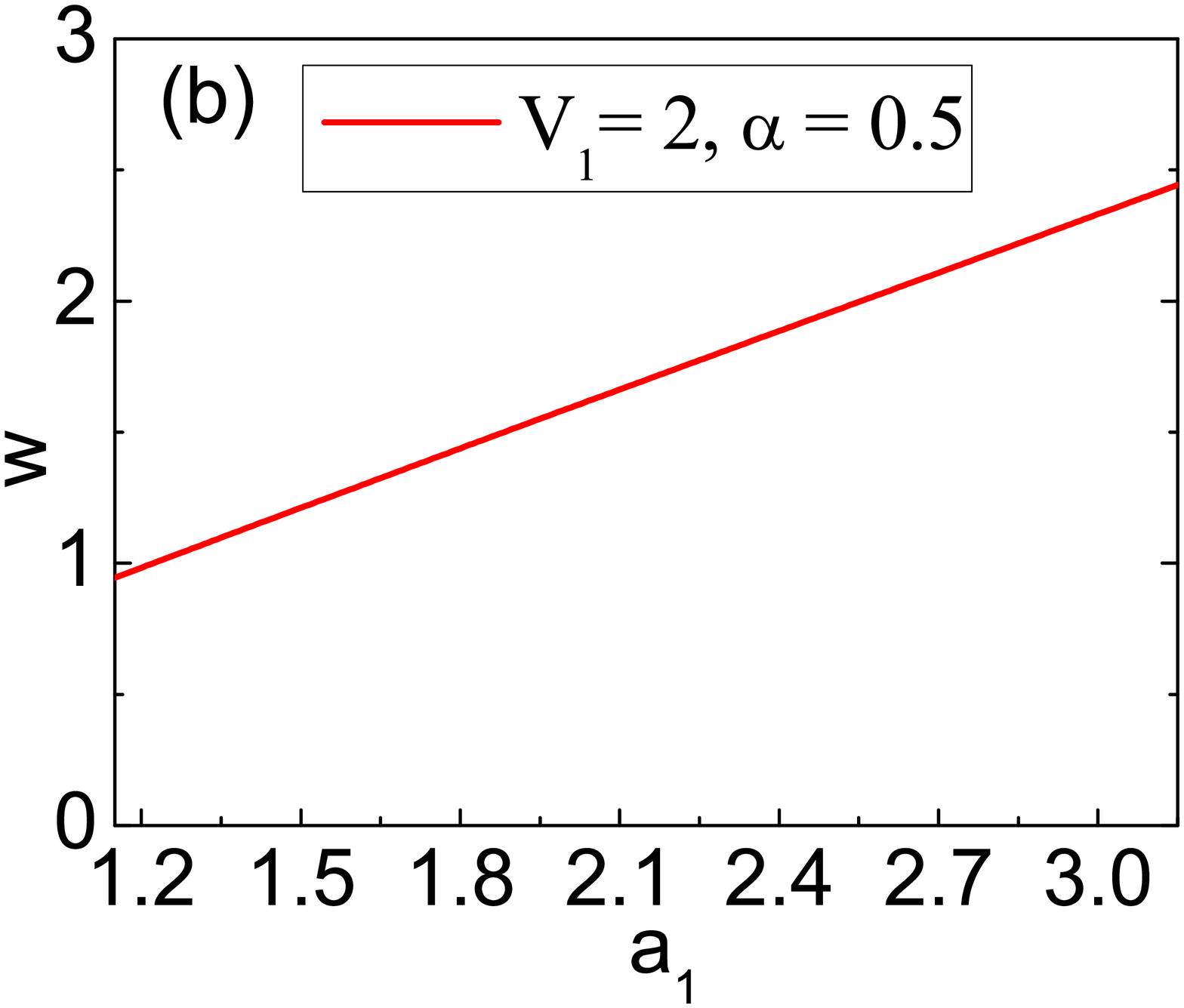}}}
\caption{ (Color online) Dip-width of stable ring-dark solitons in terms of (a) the strength $V_{1}$ and (b) width $a_{1}$ of the defect potential. The dip-width is defined from the full width at half maximal (FWHM) of the condensate density distribution in a ring-dark soliton.}
\end{figure}

After showing the stability of RDSs, we would like to investigate the properties of a stable RDS via studying how its dip-width, $w$, varies with the strength and width of the defect potential. Here we consider a low-pump power case with $\alpha=0.5$ (see Supplementary for the higher-pump power case with $\alpha=1$). In Fig. 4(a), we find that, for a constant potential width $a_{1}=2$, $w$ increases slowly as the potential strength $V_{1}$ increases. For the stronger defect potential, there is a stronger expulsion force to expel condensate polaritons away from the dip region of a RDS. This depletion leads to the decreased condensate density near the dip region. Therefore, the width of the ring soliton becomes bigger. From Fig. 4(b), we find that, for a constant potential strength $V_{1}=2$, $w$ increases more rapidly as the potential width $a_{1}$ increases. The role of the potential width here is similar to the potential strength. For the narrow cases below a critical width ($a_{1}<a_{c}\thicksim0.5$), the generated solitons are stable RGSs. There are some particles inside the dip to provide repulsive interactions for the stability of the RGSs. When the width gets broader and greater than the critical width ($a_{1}>a_{c}$), the RGSs become stable RDSs without condensate polaritons inside the dip region.

In the setting of Ref. \cite{Dominici}, the formation of the RDS is achieved by generating a drop of polariton condensate which was instantaneously created on the backdrop of a previously unperturbed state. With the time evolved, the interference phenomena of coherent waves are acting in reshaping the polariton density as a series of concentric rings around the bright peak in center. The spontaneously formed RDSs have finite lifetimes and have different structures from the RDSs discussed in this work. In the theoretical modeling of Ref. \cite{Rodrigues}, on the other hand, the RGSs are generated from a finite-size pumping configuration. Therefore, the 2D RGSs are actually an annular density depletion on top of a center-high nodeless cloud. In this article, we present a way of generating a 2D RDS with an annular intensity dip on top of a homogeneous background. The radial location of the dip can be engineered by the external potential.

In conclusion, we studied ring dark solitons occurring in microcavity-polariton condensates. We have found regions of instability and stability for the RDSs. The RDSs are stably generated when an annular defect potential with repulsion is applied. The RDSs exist stably in a regime with stronger defect-potential strengths and lower pump powers. Further investigation for the properties of the RDSs shows that the dip-width of a soliton increases with the increasing of the potential strength. For a fixed potential strength, the dip-width of a soliton also increases with the increasing of the potential width.

\bibliography{VR}

\end{document}